\documentclass[a4paper,11pt]{article}
\usepackage{jheppub}
\usepackage[T1]{fontenc}
\usepackage{youngtab}

\title{\boldmath Singlet-Assisted Supersymmetry Breaking For $Sp(2N)$ Theories}
\author[a,1]{Bin Zhu,\note{Corresponding author.}}
\author[b]{Kun Meng,}
\author[a]{Ran Ding}
\author[a]{and Qiyi Li}

\affiliation[a]{School of Physics, Nankai University,\\ China}
\affiliation[b]{School of Science, Tianjin Polytechnic University,\\China}

\emailAdd{zhubin@mail.nankai.edu.cn}
\emailAdd{dingran@mail.nankai.edu.cn}
\emailAdd{kunmeng@mail.nankai.edu.cn}
\emailAdd{liqiyi@mail.nankai.edu.cn}

\abstract{We investigate local supersymmetry-breaking vacua in s-confining theories with gauge group $Sp(2N)$. By adapting the general recipe developed by Shadmi and Shirman, we construct a realistic model based on dynamics of SQCD coupled with singlets which allows a spontaneously broken supersymmetry. Since the chiral superfields in model have R-charges $R=0$ and $R=2$ only, the tedious computations of Coleman-Weinberg potential can be greatly alleviated from the lesson of David Shih. We observe that the pseudomoduli fields are stabilized at the origin of moduli space at one-loop order with calculability being preserved.}

\begin{document}
\maketitle
\flushbottom

\section{Introduction}

Dynamical supersymmetry breaking \cite{Witten:1981kv, Witten:1981nf}
 provides compelling explanation for hierarchy problem, for reviews, see \cite{Intriligator:2007cp,Kitano:2010fa, Dine:2010cv}.
 By using dimensional transmutation, small mass scale arises naturally. The frustrating fact that models with dynamical supersymmetry breaking are non-generic can be ameliorated if we accept meta-stable supersymmetry breaking \cite{Intriligator:2006dd}. In this scenario, a weakly coupled supersymmetry breaking sector was derived from Seiberg duality \cite{Seiberg:1994pq}. The properties of supersymmetry breaking vacua are illustrated by O'Raifeartaigh models \cite{ORaifeartaigh: 1975} which are dynamically generated in IR.

The vector-like supersymmetry breaking framework including various deformations of ISS model \cite{DiMa} has been studied extensively. It is equally important to search for dynamical supersymmetry breaking vacua in chiral theories \cite{Lin:2011vd, Shadmi:2011mt, Shadmi:2011uy, Bertolini:2011hy}. In these models, s-confining theories \cite{Csaki:1996zb, Csaki:1998th} support a calculable framework for analyzing supersymmetry breaking vacua. This is mainly because s-confining theory admits a smooth description of moduli space in IR in terms of gauge invariant composites.

This paper is intended to search for dynamical supersymmetry breaking in s-confining theories with gauge group $Sp(2N)$. The basic methodology we adapt was developed by Shadmi and Shirman \cite{Shadmi:2011uy}. By introducing singlets for each gauge invariant composites $O_{i}$ except one, supersymmetry breaking vacua is easy to obtain

\begin{align}
\delta W=S_{i}O_{i}+fO.
\label{singlet}
\end{align}

Below confinement scale $\Lambda$, s-confining theory reduces to O'Raifeartaigh model with low energy degrees of freedom being gauge invariant composites. The last term in (\ref{singlet}) plays a crucial role in supersymmetry breaking as will be covered later.

In section $2$, we set up a detailed analysis on a special s-confining theory, namely $Sp(2N)$ with one anti-symmetric tensor and six fundamental flavors. We observed that it is easy to obtain supersymmetry breaking vacua in this model coupled to singlets. The component form of dynamical superpotential is

\begin{align}
W_{dyn}=\varepsilon^{abcdef}X_{ab}\phi_{cd}\phi_{ef}.
\end{align}

In section $3$, we focus our attention on the stabilization of pseudomoduli space. For single pseudomoduli, the computation of Coleman-Weinberg potential \cite{Coleman:1973jx} is compact and simple in terms of the method developed by David Shih \cite{Shih:2007av}. The non-calculable contributions are naturally suppressed due to the fact that gauge invariant composites have dimensions larger than $2$. Using naive dimensional analysis, the gauge invariant composites should be scaled by $\Lambda^{d-1}$, where $d$ is the canonical dimension of gauge invariant composites. After recaling, a small dimensionless number arises naturally

\begin{align}
\epsilon=\frac{\Lambda}{M},
\end{align}

where $M$ is the UV cut-off.

In the last section, we briefly discuss other s-confining theories coupled to singlets with gauge group $Sp(2N)$. It is different from $SU(N)$ s-confining theories that the number of s-confining theories is highly constrained, i.e., we have exhausted the studies on s-confining theories with gauge group $Sp(2N)$.

\section{Model construction}
The simplest model to break supersymmetry dynamically for $Sp(2N)$ is $Sp(4)$ gauge theory with one anti-fundamental tensor and six fundamental flavors \cite{Csaki:1996eu}. The anomaly-free symmetries and gauge invariant composites are given in Table (\ref{tab:1})

\begin{table}[htp!]
\centering
\begin{tabular}{|c|c|c|c|c|}
\hline
& $Sp(4)$ & $SU(6)$ & $U(1)$ & $U(1)_{R}$ \\
\hline
A & \tiny\yng(1,1) & 1 & -3 & 0\\
Q & \tiny\yng(1) & \tiny\yng(1) & 3 & $\frac{1}{3}$\\
$T=A^{2}$ & 1 & 1 & -6 & 0\\
$X_{0}=Q^{2}$ & 1 & \tiny\yng(1,1) & -4 & $\frac{1}{3}$\\
$X_{1}=QAQ$ & 1 & \tiny\yng(1,1) & -4 & $\frac{1}{3}$\\
\hline
\end{tabular}
\caption{\label{tab:1} The symmetries and confined spectra of the model we studied.}
\end{table}

This model admits a s-confined description in IR in terms of gauge invariant composites. Meanwhile, non-perturbative superpotential is dynamically generated so that the classical constraints are preserved

\begin{align}
W_{dyn}=\frac{1}{\Lambda^{5}}\left(\frac{1}{3}T X_{0}^{3}+\frac{1}{2}X_{0}X_{1}^{2}\right)
\end{align}

Using naive dimensional analysis, the Kahler potential is

\begin{align}
K=\frac{1}{\left|\Lambda\right|^{2}}T^{+}T+\frac{1}{\left|\Lambda\right|^{2}}X_{0}^{+}X_{0}+
\frac{1}{\left|\Lambda\right|^{4}}X_{1}^{+}X_{1}+\cdots
\end{align}

where dots denotes for higher order terms in Kahler potential that can be ignored when we study theories around the origin. In order to obtain canonical Kahler potential (canonical kinetic terms), rescale the superfields

\begin{align}
T&\rightarrow \Lambda T,\nonumber\\
X_{0}&=X\rightarrow \Lambda X,\nonumber\\
X_{1}&=\phi\rightarrow \Lambda^{2}\phi.
\label{rescale}
\end{align}

The corresponding superpotential becomes

\begin{align}
W_{dyn}=\frac{T X^3}{3 \Lambda }+\frac{1}{2}X\phi^2.
\label{super}
\end{align}

The factorization of superpotential into renormalizable and non-renormalizable parts is generic even if considering more complicated case. Retain the superpotential (\ref{super}) up to renormalizable interaction for studying behavior around the origin of moduli space only.

\begin{align}
W_{dyn}&=\frac{1}{2}\varepsilon^{abcdef}X_{ab}\phi_{cd}\phi_{ef}\nonumber\\
       &\simeq X_{56}\phi_{14}\phi_{23}-X_{46}\phi_{15}\phi_{23}+X_{45}\phi_{16}\phi_{23}\nonumber\\
       &+X_{16}\phi_{45}\phi_{23}-X_{15}\phi_{46} \phi _{23}+X_{14}\phi_{56}\phi_{23}\nonumber\\
       &-X_{56}\phi_{13}\phi_{24}+X_{36}\phi_{15}\phi_{24}-X_{45}\phi_{16}\phi_{24}\nonumber\\
       &+X_{46}\phi_{13}\phi_{25}-X_{36}\phi_{14}\phi_{25}+X_{34}\phi_{16}\phi_{25}\nonumber\\
       &-X_{45}\phi_{13}\phi_{26}+X_{35}\phi_{14}\phi_{26}-X_{34}\phi_{15}\phi_{26}\nonumber\\
       &+X_{56}\phi_{12}\phi_{34}-X_{26}\phi_{15}\phi_{34}+X_{25}\phi_{16}\phi_{34}\nonumber\\
       &+X_{16}\phi_{25}\phi_{34}-X_{15}\phi_{26}\phi_{34}-X_{46}\phi_{12}\phi_{35}\nonumber\\
       &+X_{26}\phi_{14}\phi_{35}-X_{24}\phi_{16}\phi_{35}-X_{16}\phi_{24}\phi_{35}\nonumber\\
       &+X_{14}\phi_{26}\phi_{35}+X_{45}\phi_{12}\phi_{36}-X_{25}\phi_{14}\phi_{36}\nonumber\\
       &+X_{24}\phi_{15}\phi_{36}+X_{15}\phi_{24}\phi_{36}-X_{14}\phi_{25}\phi_{36}\nonumber\\
       &+X_{36}\phi_{12}\phi_{45}-X_{26}\phi_{13}\phi_{45}+X_{23}\phi_{16}\phi_{45}\nonumber\\
       &-X_{13}\phi_{26}\phi_{45}+X_{12}\phi_{36}\phi_{45}-X_{35}\phi_{12}\phi_{46}\nonumber\\
       &+X_{25}\phi_{13}\phi_{46}-X_{23}\phi_{15}\phi_{46}+X_{13}\phi_{25}\phi_{46}\nonumber\\
       &-X_{12}\phi_{35}\phi_{46}+X_{34}\phi_{12}\phi_{56}-X_{24}\phi_{13}\phi_{56}\nonumber\\
       &+X_{23}\phi_{14}\phi_{56}-X_{13}\phi_{24}\phi_{56}+X_{12}\phi_{34}\phi_{56} .
       \label{Yukawa}
\end{align}\footnote{$\simeq$ means that we absorb the coefficient $8$ for it being irrelevant.}

We would like to deform this model in a way that results in supersymmetry breaking triggered by singlets.

\begin{align}
\delta W=S_{i}O_{i}+fO.
\label{singlet}
\end{align}

The values of $f$ strongly affects the behavior of dynamics near the origin. In order for gauge mediation \cite{Dine:1994vc, Dine:1995ag}, we would like to preserve large part of the global symmetry group $SU(6)$. The simplest choice is that only $f_{12}$ is non-zero. For convenience, we single out $X_{12}$\footnote{In the following, we write it $X$ for short.} from other $X_{ab}$ and call it $X$, and the rest of pseudomoduli are called $Y$. Then the schematic version of superpotential can be written as

\begin{align}
W=\frac{1}{2}(X+Y)\phi^2
\end{align}

The dangerous flat direction $T$ can be lifted by singlets \cite{Shadmi:2011uy}. Under the same rescaling of fields (\ref{rescale}), the singlet perturbation becomes

\begin{align}
\delta W=f \Lambda X+\frac{m}{M}\varphi\phi+mST\rightarrow f\Lambda^2 X+m\frac{\Lambda}{M}\Lambda\varphi\phi+m\Lambda ST,
\end{align}

where $S, \varphi$ are additional singlets to lift dangerous flat directions. For convenience, absorb $\Lambda$ and $\epsilon$ into new parameters,

\begin{align}
f\Lambda^{2}&\rightarrow f,\\
m\Lambda\epsilon&\rightarrow m,\\
m\Lambda&\rightarrow m_{h}.
\label{mass2}
\end{align}

We observed that $m_{h}\gg m$ from (\ref{mass2}). Near the origin of moduli space, we can simplify the superpotential by integrating out the heaviest fields $S$ and $T$. Then effective O'Raifeartaigh model becomes

\begin{align}
W=fX+\frac{1}{2}(X+Y)\phi^{2}+m\phi\varphi
\end{align}

This type of superpotential is equivalent to the one studied by Curtin et.al.\cite{Curtin:2012yu}. In this scenario, they show that the mass correction of pseudomoduli $X$ and $Y$ is semi-positive definite. Some pseudomoduli remains massless at one-loop level. This raises the possibility to break R-symmetry spontaneously in two or higher loops effective potential \cite{Amariti:2008uz}. The remaining massless pseudomoduli at one-loop level is disfavored, because the two-loop corretions might be negative which destabilizes the potential around the origin and leads to runaway in general. When introducing additional tree level term $f_{y} Y$, this type of model can be remedied. Such terms give rise to a positive mass for $Y$ automatically. The tension between one-loop and two-loop contribution can be used to break R-symmetry from the renormalization group flow
\cite{Amariti:2012pg}.

In this paper, the extra pseudomoduli $Y$ can be lifted by additional singlets. The singlet perturbation is thus

\begin{align}
\delta W=fX+m\varphi\phi +m_{h} S_{Y}Y
\end{align}

Recall that $m_{h}\gg m$, $S_{Y}$ and $Y$ can be integrated out which greatly simplifies the cubic term (\ref{Yukawa}),

\begin{align}
W_{cubic}=X \phi _{36} \phi _{45}-X \phi _{35} \phi _{46}+X
   \phi _{34} \phi _{56}.
   \label{SYUkwa}
\end{align}

In addition, we mention that only $m (\phi _{36}\varphi_{36}+\phi _{45}\varphi_{45}+\phi _{35}\varphi_{35}+\phi _{46}\varphi_{46}+\phi _{34}\varphi_{34}+\phi _{56}\varphi_{56})$ are relevant for computing Coleman-Weinberg potential. Other mass terms without coupling to $X$ just give rise to constant contribution for Coleman-Weinberg potential. Therefore, the effective O'Raifeartaigh model near the origin is

\begin{align}
W=f X+m\sum_{i=1}^{6}\phi_{i}\varphi_{i}+X( \phi
   _1 \phi _2- \phi _3 \phi
   _4+\phi _5\phi _6)\\
   \label{final}
\end{align}

with

\begin{align}
\phi _{36}\to \phi _1,\phi _{45}\to \phi _2,\phi
   _{46}\to \phi _3,\phi _{35}\to \phi _4,\phi _{34}\to
   \phi _5,\phi _{56}\to \phi _6,\nonumber\\ \varphi_{36}\to \varphi
   _1,\varphi_{45}\to \varphi _2,\varphi_{46}\to \varphi _3,\varphi_{35}\to
   \varphi _4,\varphi_{34}\to \varphi _5,\varphi_{56}\to \varphi
   _6.
\end{align}

The F-term of this model reads,

\begin{align}
\bar F_{X}&=f+\frac{h}{2}\phi^2\nonumber,\\
\bar F_{\varphi}&=m\phi\\
\bar F_{\phi}&=hX\phi +m\varphi\nonumber
\end{align}

Obviously, the first two equations are not compatible, thus result in spontaneous supersymmetry breaking. Before falling into actual computation on the quantum effective potential for pseudomoduli, we mention that $f$ must be much smaller than $m^2$ in order to have the pseudomoduli space.

\begin{align}
X \quad arbitrary, \quad  \phi_{i}=\varphi_{i}=0.
\end{align}

This inevitable hierarchy can not be generated naturally as (\ref{mass2}), and has to be imposed by hand.

\section{Coleman-Weinberg potential}

In this section, we tried to compute the mass of pseudomoduli $X$ in the context of the general formula developed by David Shih \cite{Shih:2007av}. Our model (\ref{final}) is the conventional O'Raifeartaigh model with R charge equaling to $0$ or $2$.

\begin{align}
R(X)&=2,\nonumber\\
R(\phi)&=0,\nonumber\\
R(\varphi)&=2.
\label{R-charge}
\end{align}

Since superfields have R-charge $0$ or $2$ only, we concluded that this model has a one-loop positive-definite mass \cite{Shih:2007av}
 , thus the pseudomoduli $X$ is stabilized at the origin. Delicate computations convinced us that the supersymmetry breaking vacuum at origin is stable.

First of all, the mass matrix $M$ is thus
\begin{align}
M=\left(
\begin{array}{cccccccccccc}
 0 & 0 & 0 & 0 & 0 & 0 & m & 0 & 0 & 0
   & 0 & 0 \\
 0 & 0 & 0 & 0 & 0 & 0 & 0 & m & 0 & 0
   & 0 & 0 \\
 0 & 0 & 0 & 0 & 0 & 0 & 0 & 0 & m & 0
   & 0 & 0 \\
 0 & 0 & 0 & 0 & 0 & 0 & 0 & 0 & 0 & m
   & 0 & 0 \\
 0 & 0 & 0 & 0 & 0 & 0 & 0 & 0 & 0 & 0
   & m & 0 \\
 0 & 0 & 0 & 0 & 0 & 0 & 0 & 0 & 0 & 0
   & 0 & m \\
 m & 0 & 0 & 0 & 0 & 0 & 0 & 0 & 0 & 0
   & 0 & 0 \\
 0 & m & 0 & 0 & 0 & 0 & 0 & 0 & 0 & 0
   & 0 & 0 \\
 0 & 0 & m & 0 & 0 & 0 & 0 & 0 & 0 & 0
   & 0 & 0 \\
 0 & 0 & 0 & m & 0 & 0 & 0 & 0 & 0 & 0
   & 0 & 0 \\
 0 & 0 & 0 & 0 & m & 0 & 0 & 0 & 0 & 0
   & 0 & 0 \\
 0 & 0 & 0 & 0 & 0 & m & 0 & 0 & 0 & 0
   & 0 & 0
\end{array}
\right)
\label{M}
\end{align}

and $N$ is

\begin{align}
N=\left(
\begin{array}{cccccccccccc}
 0 & 1 & 0 & 0 & 0 & 0 & 0 & 0 & 0 & 0
   & 0 & 0 \\
 1 & 0 & 0 & 0 & 0 & 0 & 0 & 0 & 0 & 0
   & 0 & 0 \\
 0 & 0 & 0 & -1 & 0 & 0 & 0 & 0 & 0 & 0
   & 0 & 0 \\
 0 & 0 & -1 & 0 & 0 & 0 & 0 & 0 & 0 & 0
   & 0 & 0 \\
 0 & 0 & 0 & 0 & 0 & 1 & 0 & 0 & 0 & 0
   & 0 & 0 \\
 0 & 0 & 0 & 0 & 1 & 0 & 0 & 0 & 0 & 0
   & 0 & 0 \\
 0 & 0 & 0 & 0 & 0 & 0 & 0 & 0 & 0 & 0
   & 0 & 0 \\
 0 & 0 & 0 & 0 & 0 & 0 & 0 & 0 & 0 & 0
   & 0 & 0 \\
 0 & 0 & 0 & 0 & 0 & 0 & 0 & 0 & 0 & 0
   & 0 & 0 \\
 0 & 0 & 0 & 0 & 0 & 0 & 0 & 0 & 0 & 0
   & 0 & 0 \\
 0 & 0 & 0 & 0 & 0 & 0 & 0 & 0 & 0 & 0
   & 0 & 0 \\
 0 & 0 & 0 & 0 & 0 & 0 & 0 & 0 & 0 & 0
   & 0 & 0
\end{array}
\right)
\label{N}
\end{align}

From (\ref{M}) and (\ref{N}), we can obtain the important element $\mathcal{F}$ that is proportional to $f/(m^{2}+v^{2})$\footnote{The matrix of $\mathcal{F}$ is too large to display}. Recall that there is a general mass formula for mass squared

\begin{align}
m_{X}^{2}=M_{1}^{2}-M_{2}^{2},
\end{align}

where

\begin{align}
M_{1}^{2}&=\frac{1}{16\pi^{2}f^{2}}\int_{0}^{\infty}dv v^{5} Tr\left[\frac{\mathcal{F}(v)^{4}}{1-\mathcal{F}(v)^2}\right]£¬\\
M_{2}^{2}&=\frac{1}{8\pi^{2}f^{2}}\int_{0}^{\infty}dv v^{3} Tr\left[\left(\frac{\mathcal{F}(v)^{2}}{1-\mathcal{F}(v)^2}\hat M\right)^{2}\right]¡£
\end{align}

and

\begin{align}
\hat M&=\left(\begin{array}{cccccccccccc}
 0 & M^{+}  \\
 M & 0
 \end{array}
 \right)\\
\hat N&=\left(\begin{array}{cccccccccccc}
 0 & N^{+}  \\
 N & 0
 \end{array}
 \right)\\
 \mathcal{F}&=(v^2+\hat M^{2})^{-1/2}f\hat N(v^2+\hat M^2)^{-1/2}
\end{align}

 Straightforward calculation shows that $M_{2}^{2}$ is identically zero for this model. The first term $M_{1}^{2}$ at small $y=f/m^2$ gives

\begin{align}
m_{X}^{2}=\frac{m^6 y^4}{8 \pi ^2 f^2}
\end{align}

Thus, we obtain the positive mass for $X$ with the metastable vacuum located at the origin of moduli space.

Another issue to be addressed is the calculability of this model. Recall that non-calculable contributions come from either non-renormalizable terms in superpotential or higher order operators in Kahler potential. Non-renormalizable terms in superpotential are indeed negligible for all vevs being much smaller than $\Lambda$.  We thus focus our attentions on the Kahler potential

\begin{align}
K=X X^{+} \left(1+g_{X}\left(\frac{X}{\Lambda},\cdots\right)\right).
\end{align}

The potential near the origin is

\begin{align}
V=f^2 \left(\frac{\alpha  X^2}{\Lambda ^2}+1\right)
\end{align}

so non-calculable contribution to mass is

\begin{align}
\delta m_{non-calc}^2=\frac{f^2}{\Lambda ^2}
\end{align}

The ratio between non-calculable contributions and calculable one is naturally small

\begin{align}
\gamma\equiv\frac{\delta m_{non-calc}^2}{\delta m_{calc}^{2}}=\frac{24 \pi ^2m^2\epsilon^2}{
   \Lambda^2}=24\pi^2\epsilon ^4\sim\epsilon ^4
\end{align}

The condition $\gamma\ll 1$ is highly non-trivial, since it indicates that the non-calculable corrections is suppressed by the presence of small parameter $\epsilon$. The model under consideration has calculable dynamical supersymmetry breaking vacuum.

\section{Other s-confining theories}

There are only two types of s-confining theories with gauge group $Sp(2N)$ which can be seen in \cite{Csaki:1996eu}. The complicated one has been discussed in detail in the last two sections. Here we concentrate on the simpler one, i.e., $Sp(2N)$ with $2N+4$ fundamental flavors. The matter content and global symmetries can be found in Table (\ref{tab:i})

\begin{table}[htp!]
\centering
\begin{tabular}{|c|c|c|c|c|}
\hline
 & $Sp(2N)$ & $SU(2N+4)$ & $U(1)_{R}$  \\
\hline
Q & $\tiny\yng(1)$ &$\tiny\yng(1)$ & $\frac{1}{N+2}$  \\
\hline
$X=Q^{2}$ & $1$ & $\tiny\yng(1,1)$ & $\frac{1}{N+2}$  \\
\hline
\end{tabular}
\caption{\label{tab:i} The matter content and global symmetries of new model.}
\end{table}

The dynamical superpotential is $W=\Lambda^{1-N}X^{N+2}$. At IR, no renormalizable interactions are permitted. Any perturbation can break supersymmetry. However, this scenario is non-calculable.

\section{Conclusion}

S-confining theories provide excellent avenues in which searching for dynamical supersymmetry breaking is available. By coupling singlets to the theory, it exhibits well-behaved characters in IR, for example, the calculability can be preserved. 

In our first model, namely $Sp(4)$ gauge theory with one anti-fundamental tensor and six fundamental flavors, the dynamics in IR has been analyzed extensively. The positive definite mass correction shows that it does have stable supersymmetry breaking vacuum at the origin of moduli space. From the perspective of calculability, we found that it is guaranteed naturally without imposing hierarchy between couplings by hand. Since the fields in this model have R-charge $0$ or $2$ only, R-symmetry can not be broken at one-loop quantum corrections. The slightly deformed model may induce R-symmetry breaking which is left for future study.

Another model, $Sp(2N)$ with $2N+4$ fundamental flavors, can break supersymmetry easily. Unfortunately, this scenario is non-calculable.

\end{document}